\documentclass[
  aps,
  prl,
  reprint,
  showpacs,
  floatfix,
  superscriptaddress,
  footinbib
]{revtex4-2}

\usepackage[T1]{fontenc}
\usepackage[utf8]{inputenc}
\usepackage[english]{babel}
\usepackage[normalem]{ulem}

\usepackage{amsmath}
\usepackage{mathrsfs} 
\usepackage{lipsum}
\usepackage{amsfonts}
\usepackage{amssymb}
\usepackage{amsthm}
\usepackage{mathtools}
\usepackage{latexsym}
\usepackage{bm}
\usepackage{relsize}
\usepackage{braket}
\usepackage{slashed}
\usepackage{empheq}
\usepackage[makeroom]{cancel}
\usepackage{xfrac} % slanted fractions
\usepackage{upgreek}

\usepackage{graphicx}
\usepackage{tabularx}
\usepackage{multirow}
\usepackage{floatrow}

\usepackage{pgf}

\usepackage{xspace}
\usepackage{hyperref} 
\usepackage[most]{tcolorbox}
\usepackage{comment}

\usepackage{tikz-feynman}
\usepackage{tikz}	
\tikzfeynmanset{compat=1.0.0}

\allowdisplaybreaks

\usepackage{colortbl}
\definecolor{summersky}{cmyk}{0.71,0.33,0,0.5}
\definecolor{flamingo}{cmyk}{0,0.51,0.71,0.5}
\definecolor{rp}{cmyk}{0.2, 1, 0.6, 0}
\definecolor{pacificblue}{cmyk}{0.95,0.3,0, 0.5}
\definecolor{gray60}{cmyk}{0.4,0.4,0,0.8}

\newcolumntype{P}[1]{>{\centering\arraybackslash}p{#1}}

\newcommand{\bfx}{{\mathbf{x}}}
\newcommand{\bfk}{{\mathbf{k}}}

\newcommand{\Mpl}{M_{\mathrm Pl}}

\newcommand{\dd}{\mathrm{d}}

\begin{document}

\title{Phenomenology of an Open Effective Field Theory of Dark Energy}

 \author{Santiago Ag\"u\'i Salcedo}
 \affiliation{Kavli Institute for Cosmological Physics,
University of Chicago, 5640 South Ellis Ave.,
Chicago, IL 60637, USA}
\affiliation{Department of Applied Mathematics and Theoretical Physics, University of Cambridge, Wilberforce
Road, Cambridge, CB3 0WA, UK}
\author{Thomas Colas}
 \affiliation{Department of Applied Mathematics and Theoretical Physics, University of Cambridge, Wilberforce
Road, Cambridge, CB3 0WA, UK}
\author{Lennard Dufner}
 \affiliation{Department of Applied Mathematics and Theoretical Physics, University of Cambridge,
Wilberforce Road, Cambridge, CB3 0WA, UK}
\author{Enrico Pajer}
 \affiliation{Department of Applied Mathematics and Theoretical Physics, University of Cambridge,
Wilberforce Road, Cambridge, CB3 0WA, UK}

\begin{abstract}

All observational evidence for dark matter and dark energy is so far exclusively gravitational. Hence, the dark sector may be equivalently described by a theory of the spacetime metric whose dynamics is affected by interactions with an unknown environment. Adapting open-system techniques, we have recently constructed such a general theory of open gravitational dynamics. Here we study a minimal and concrete realization of this theory that describes the late-time acceleration of the Universe. 
Our model provides a good fit to recent baryon acoustic oscillation measurements by construction, while avoiding violations of the null energy condition. 
Moreover, it leads to a set of correlated and observationally testable predictions. Studying the modified cosmological perturbation theory and compared to the $\Lambda$CDM model we find: a dissipative suppression of the gravitational-wave luminosity distance relative to the electromagnetic one; a modification in the evolution of the Bardeen potentials with a clear signal in the gravitational slip; and an enhancement of structure formation at low redshift. We present semi-analytical estimates of the magnitude of these effects and show that they lie within the reach of current constraints while providing clear targets for upcoming cosmological surveys.

\end{abstract}

\maketitle
 
\paragraph*{Introduction.---}
Measurements of Baryon Acoustic Oscillations (BAO) in DESI DR2 have provided tantalizing hints that dark energy may be a dynamical entity \cite{DESI:2025zgx}. While these results are still preliminary and require confirmation by independent probes, they have already stimulated renewed interest in departures from the standard cosmological constant paradigm. If the BAO measurements are interpreted assuming that dark matter and dark energy are separately conserved, the data show a parameterization-independent preference for models in which dark energy violates the Null Energy Condition  \cite{DESI:2025fii,Lewis:2024cqj,Figueruelo:2026eis}. Such violations are often associated with gradient or ghost instabilities in many field-theoretic descriptions. If taken at face value, this tension provides a strong motivation to explore frameworks in which the dark sector is not composed of independently conserved components, but instead allows for interactions between dark matter and dark energy \cite{Amendola:1999er, Guo:2004vg, Cai:2004dk, McDonough:2021pdg,Bedroya:2025fwh, Khoury:2025txd}. 

At the same time, recent theoretical developments have introduced a set of open-system techniques for describing gravitational dynamics in the presence of an unknown medium \cite{Salcedo:2024smn,  Lau:2024mqm, Salcedo:2025ezu, Christodoulidis:2025vxz}. The goal of this paper is to illustrate, through a concrete model, how this open-system perspective can reproduce BAO measurements, respects the null-energy condition and predicts a set of correlated signatures in cosmological perturbations and gravitational waves, which can be tested with current and forthcoming observations.

\paragraph*{Open Effective Field Theories.} 
The open system approach to gravity is an active and rapidly developing line of research
\cite{Hongo:2018ant, Salcedo:2024smn, Salcedo:2024nex, Lau:2024mqm, Salcedo:2025ezu, Christodoulidis:2025ymc, Christodoulidis:2025vxz, Colas:2025app, Ha:2026kie}. In \cite{Salcedo:2025ezu}, we built a class of theories, based on general relativity and the Schwinger-Keldysh formalism \cite{Schwinger:1960qe, Keldysh:1964ud}, to study gravity in the presence of an unknown medium. This open theory of gravity provides a systemic description of dissipative and open system gravitational dynamics. 
A few foundational aspects are still under investigation, including a prescription to count propagating degrees of freedom, additional consistency conditions on dissipative operators \cite{Christodoulidis:2025vxz}, and the non-linear realization of gauge symmetries \cite{Kaplanek:2025moq}.
In this work, we set aside these interesting open questions and instead concentrate on a minimal model, whose internal consistency has been established in \cite{Christodoulidis:2025vxz}. We use this as poster child of the rich phenomenology that emerges in the large class of open models of the dark sector. 

More concretely, the open system we wish describe are metric fluctuations and the environment is the dark sector. The addition of baryons will be discussed later. The generic local theory of open gravity is constructed in unitary gauge from a Schwinger-Keldysh functional of the form
\begin{align}
    S_{\mathrm{eff}} &= \int \dd^4 x \sqrt{-g} \Big[M_{\mu\nu} (g_{\mu\nu}, g^{00}, K_{\mu\nu}, \nabla_\mu; t) a^{\mu\nu} \nonumber
    %\nonumber \\
    % +& i N_{\mu\nu\rho\sigma} (R_{\mu\nu\rho\sigma}, g^{00}, K_{\mu\nu}, \nabla_\mu; t) a^{\mu\nu}  a^{\rho\sigma} + 
    + \cdots \Big] . 
\end{align}
Here $a^{\mu\nu}$ is an auxiliary field, known as the advanced (inverse) metric and $M_{\mu\nu}$ is a rank-$2$ co-tensor under retarded time-dependent spatial diffs, which generically breaks retarded time diffs. Building on the seminal work of the EFT of Inflation \cite{Cheung:2007st} and Dark Energy \cite{Gubitosi:2012hu}, this tensor is constructed out of the metric $g_{\mu\nu}$, a unit co-vector $n_\mu$, which in this gauge reads $n_{\mu} = - \delta^0_{~\mu}/\sqrt{-g^{00}}$, and the extrinsic curvature of the associated constant-time foliation
$K_{\mu\nu} = \left(\delta_{~\mu}^{\sigma} + n_\mu n^\sigma \right) \nabla_\sigma n_\nu$. The ellipsis in $S_{\mathrm{eff}}$ represent higher orders in the advanced metric $a^{\mu\nu}$ that do not contribute to the deterministic dynamics of the theory, on which we focus. 

Let us briefly comment on the structure underlying this theory. The fields $g_{\mu\nu}$ and $a^{\mu\nu}$ arise as the symmetric and anti-symmetric combinations of the metric field evaluated on the two branches of Schwinger-Keldysh path integral. 
These two metrics come endowed with two copies of diff invariance, one in each branch. In the presence of dissipative effects, the technical implications of these symmetries depend on one's perspective and choice of variables. In \cite{Salcedo:2024nex,Christodoulidis:2025ymc,Salcedo:2025ezu} the emergence of a deformed advanced symmetry was observed, which acts simply on the natural low energy variables, namely $g_{\mu\nu}$ and $a^{\mu\nu}$ in our case. To linear order, this symmetry always arises whenever a retarded symmetry is imposed, however at non-linear order things remained unclear. The nice work in \cite{Kaplanek:2025moq} demonstrated for quantum electrodynamics that two gauge symmetries that act in the standard way on the natural fields of the high-energy theory must always survive as a consequence of a diagonal BRST symmetry. This has important implication for the general structure of open gravity, which we hope to explore in the future. However, this does not affect the linear-order phenomenology we study in this paper and so we don't discuss it further. Moreover, \cite{Christodoulidis:2025vxz} pointed out that only some specific choices of dissipative operators in $M_{\mu\nu}$ lead to a consistent dynamics. Here we focus on a particular choice of dissipation that was shown to be consistent in \cite{Christodoulidis:2025vxz}, postponing a systematic investigation to future work. 

Our minimal model comprises of the universal sector of the EFT of Inflation \cite{Cheung:2007st} and Dark Energy \cite{Gubitosi:2012hu}, supplemented by a speed-of-sound operator and a consistent dissipative operator at first order in derivatives \cite{Christodoulidis:2025vxz}. Hence, at linear order in $a^{\mu\nu}$ our model is described by the open functional
\begin{align}%\label{eq:action_cons_operator}
    &S_{\mathrm{eff}}  = \int \dd^{4}x\sqrt{-g}\bigg(\bigg\{\Mpl^{2}\left[G_{\mu\nu}+ \Gamma(t) \kappa_{\mu\nu}\right] + \big[\Lambda(t) \nonumber \\
     +& c(t)g^{00} \big] g_{\mu\nu}\bigg\} \frac{a^{\mu\nu}}{2}  -\left[ c(t) - \frac{\Mpl^{2}}{2} M^2(t) (1 + g^{00}) \right] a^{00} \bigg) \nonumber , 
\end{align}
where $\kappa_{\mu\nu} = \sqrt{-g^{00}} \left[ K_{\mu\nu} - K (g_{\mu\nu} + n_\mu n_\nu) \right]$ and $\Gamma(t)$, $c(t)$, $\Lambda(t)$ and $M(t)$ are unknown functions of time. The contribution $\Gamma(t) \kappa_{\mu\nu}$ encodes genuinely dissipative dynamics and a departure from the familiar closed EFT of cosmological perturbations. At the classical, deterministic level, all the predictions of this model follow from the modified Einstein equations, obtained by varying $S_{\mathrm{eff}}$ with respect to $a^{\mu\nu}$, 
\begin{align}\label{eq:EE}
    &\frac{\Mpl^{2}}{2}\left[G_{\mu\nu}+ \Gamma(t) \kappa_{\mu\nu}\right] + \frac{1}{2}\left[\Lambda(t) +c(t)g^{00} \right] g_{\mu\nu} \nonumber \\
    &\quad - \left[c(t) - \frac{\Mpl^{2}}{2} M^2(t) (1 + g^{00}) \right] \delta^0_{~\mu} \delta^0_{~\nu} = 0\,.
\end{align}
We study these equations for a homogeneous and isotropic background first, and then move to linear inhomogeneous perturbations. Our model aims to describe the late cosmic acceleration and we will always assume $\Gamma(t)$ vanishes at early times as $t\to 0$, so that physics at high redshift, say $z\gg 10$, including the CMB, is unchanged compared to the cold dark matter (CDM) model. We treat baryons as a probe and discuss them around \eqref{eq:growth_of_structure}, where we study their power spectrum.

\paragraph*{Expansion and Baryonic Acoustic Oscillations.---} 

\begin{figure}[t]
    \includegraphics[width=1\linewidth]{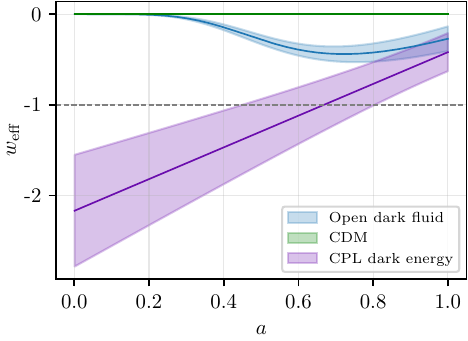}
    \caption{Evolution of effective equation of state $w_{\mathrm{eff}}$ in our model (ODF) compared that of CDM and dark energy in the CPL-parametrization. Parameters are set to the $w_0 w_a$CDM best-fit values from BAO + CMB data, with $1\sigma$-uncertainties shown \cite{DESI:2025zgx}.
    }
    \label{fig:w_a_UDF}
\end{figure}

Since we modify Einstein equations already at the background level, a first constraint on our model comes from measurements of the expansion history, as inferred from BAO and the CMB angular scale. Throughout we use BAO data from the DESI collaboration \cite{DESI:2025zgx} and CMB data from Planck \cite{Planck:2018vyg}. We do not consider supernovae data. Evaluating the Einstein equations \eqref{eq:EE} on a Friedmann-Lemaître-Robertson-Walker (FLRW) background $\dd s^2 = - \dd t^2 + a^2(t) \dd^2 \bfx$ and identifying $c = (\rho + p)/2$ and $\Lambda = (\rho - p)/2$, we obtain the background evolution. The first Friedmann equation remains standard, whereas the second Friedmann equation is modified:
\begin{align}
    3 \Mpl^2 H^2 &= \rho_{},  &   \Mpl^2 \dot{H} &= -\frac{ \rho_{} + p_{} }{2} -   \Mpl^2 \Gamma H . \label{eq:F2}
\end{align}
Cosmic acceleration, $\ddot a > 0$, occurs provided
$p/\rho < -1/3 - 2\Gamma/(3H)$, In particular, late-time acceleration can be sustained even for $p \simeq 0$, if $\Gamma < -H/2$. Combining Friedmann's equations we derive a modified continuity equation for the full system, open system plus environment:
\begin{equation}\label{eq:continuity}
    \dot{\rho}_{} + 3 H \left( \rho_{} + p_{} \right) = - 2 \Gamma \rho_{}\,.
\end{equation}
Note that $T_{\mu\nu}$ obeys a non-standard conservation equation, reflecting the intrinsically dissipative nature of the theory. The $\Gamma$ term is reminiscent of bulk viscosity and for $\Gamma < 0$, i.e.~anti-viscosity, the fluid sustains late-time cosmic acceleration as already discussed in \cite{Zimdahl:2000zm}. Solving \eqref{eq:continuity} for $p=0$ yields
\begin{equation}\label{eq:rhoodf}
    \rho = \frac{\rho_{0}}{a^3} \exp \left[ -\int_1^a \dd \tilde{a} \, \frac{2 \Gamma(\tilde{a})}{\tilde{a} H(\tilde{a})} \right] .
\end{equation}
We know that that the evolution of CDM plus a decoupled dark energy with CPL parametrization \cite{Chevallier:2000qy,Linder:2002et},
\begin{align}\label{eq:w0wa}
     \rho = \frac{\rho_{0}}{a^3} \left[ \Omega_{\mathrm{m},0} + (1 - \Omega_{\mathrm{m},0}) a^{-3(w_0+w_a)} e^{3 w_a (a-1)} \right] \,.
\end{align}
is a good fit to DESI. We therefore use it to derive a simple analytical expression for $\Gamma(t)$ that fits equally well. Equating \eqref{eq:w0wa} to \eqref{eq:rhoodf} gives
\begin{equation}\label{eq:gamma}
    \Gamma =  \frac{3H}{2}  \frac{\left[w_0 + w_a(1-a)\right] \left(1 -\Omega_{\mathrm{m},0}\right)}{\left(1 -\Omega_{\mathrm{m},0} \right)+ \Omega_{\mathrm{m},0}  a^{3(w_0+w_a)}e^{3 w_a (1-a)}} .
\end{equation}
With this choice, our model of a dissipative unified dark fluid fits well DESI DR2 (see \cite{Kou:2025yfr} for a very similar construction). Comparing to the standard continuity equation, an effective equation of state parameter  can be defined by $w_{\mathrm{eff}}\equiv-\dot \rho/(3H\rho)-1$. For our model, which we dub Open Dark Fluid (ODF) this is $w_{\mathrm{eff}} = 2\Gamma/(3H)$, which is shown in Fig. \ref{fig:w_a_UDF}. Notably, $w_{\mathrm{eff}}$ never crosses the phantom divide, $w_{\mathrm{eff}} = -1$, even though such a crossing may appear when the dark sector is artificially decomposed into separately conserved cold dark matter and dark energy components. This is a well known feature \cite{Das:2005yj, Khoury:2025txd, Kou:2025yfr, Chakraborty:2025syu} of interacting dark sector models. \\ 

\paragraph*{Gravitational wave luminosity distance.---} 
A key virtue of our model is that it yields consistent predictions for both the background and perturbations, enabling a correlated analysis of their effects. To illustrate this point, we consider transverse-traceless perturbations defined by $g_{ij} = a^2(t)\left(\delta_{ij} + \gamma_{ij}\right)$. Using the expression for the transverse-traceless component $\left( \delta \kappa_{ij} \right)_{\rm TT} = a^2/2 \dot{\gamma}_{ij} - 2 a^2 H \gamma_{ij}$, we find a modified equation for gravitational waves
\begin{equation}
    \ddot{\gamma}_{ij} + (3 H + \Gamma) \dot{\gamma}_{ij} - \frac{\nabla^2}{a^2} \gamma_{ij} = 0 \,.
\end{equation}
The dissipative term controlled by $\Gamma$ induces an additional friction term in the gravitational-wave (GW) equation of motion \cite{Lau:2024mqm,Salcedo:2025ezu}. For $\Gamma > 0$, GWs experience an amplitude suppression during propagation, in contrast to vacuum propagation in General Relativity, thereby encoding the imprint of the medium. Analogous effects from a different origin were observed in the closed EFT of Dark Energy \cite{Creminelli:2018xsv,Creminelli:2019nok,Creminelli:2019kjy}. As a result of $\Gamma$, sources appear more distant if their propagation is interpreted as conservative. The opposite behavior arises for $\Gamma < 0$. This effect can be quantified by comparing the GW luminosity distance, $d_L^{\rm gw}(z)$, to its electromagnetic counterpart, $d_L^{\rm em}(z)$, through \cite{Belgacem:2018lbp}
\begin{align}\label{eq:GW_lum_dist_pheno}
\frac{d_L^{\,\rm gw}(z)}{d_L^{\,\rm em}(z)} &= \exp \left[ \frac{1}{2} \int_0^z \frac{\dd \tilde{z}}{(1+\tilde{z})H(\tilde{z})} \Gamma(\tilde{z}) \right] \,.
\end{align}

\begin{figure}[t]
    \includegraphics[width=1\linewidth]{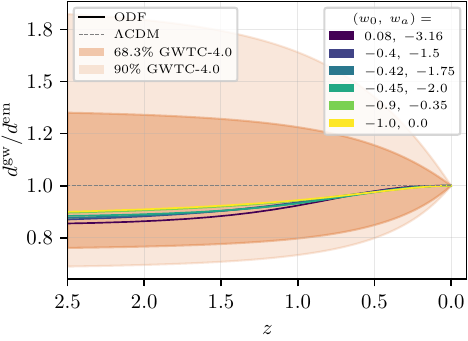}
    \caption{Gravitational wave luminosity distance predicted by our model (ODF) for $(w_0, w_a)$ values within the BAO+CMB 95\% confidence region, versus GWTC-4 constraints. Shaded regions show $68.3\%$ ($90\%$) credible intervals from GWTC-4 fits to the $\alpha_m$-parametrization, assuming a narrow $H_0$-prior \cite{LIGOScientific:2025jau}.}
    \label{fig:GWlum}
\end{figure}

Substituting the expression for $\Gamma$ in \eqref{eq:gamma}, we evaluate $d_L^{\rm gw}/d_L^{\rm em}$ for a set of representative $(w_0, w_a)$-points chosen from within the BAO+CMB $2\sigma$-ellipse \cite{DESI:2025zgx}, shown as individual curves in Fig.~\ref{fig:GWlum}. Superimposing the bounds from the LIGO–Virgo–KAGRA Collaboration Gravitational-Wave Transient Catalog (GWTC-4.0) \cite{LIGOScientific:2025jau}, shown in orange, we find that our model's prediction for the gravitational-wave luminosity distance is consistent with current observational constraints. Improved gravitational-wave measurements will provide a decisive test. The general caveats about using the same EFT for cosmological and interferometer scales also apply to our model \cite{deRham:2018red}. \\

\paragraph*{Galaxy clustering and weak lensing.---}

Our model leaves distinct imprints on the growth of structure and the bending of light. These effects are captured by the linear evolution of scalar perturbations, encoded in the Bardeen potentials. We parametrize the scalar sector of the perturbed metric following Weinberg's notations as
\begin{equation}\label{eq:perturbed_metric_perseas}
    \delta g_{\mu\nu}=\left(
    \begin{array}{c@{\hspace{0.8em}}c}
     -E & a (t)\partial_i F\qquad  \\[0.8em]
     a (t)\partial_i F & a^2(t)\left[A \delta_{ij} + \partial_i \partial_j B \right] \\
    \end{array}
    \right).
\end{equation}
It is convenient to introduce the combination $\sigma \equiv a\left(1/2 \, a \dot{B} - F\right)$, which is invariant under spatial diffs. Consequently, any equation involving only $E$, $A$, and $\sigma$ is automatically free of spatial gauge redundancies.

The linearized Einstein equations take the form \cite{Christodoulidis:2025vxz}
\begin{align}
    - 2\frac{\partial^2}{a^2} A + 6 H \dot{A} + 4 H \frac{\partial^2}{a^2} \sigma - 3 H^2 E   + 2 M^2 E  &= 0 ,  \label{eq:00v2} \\
    H E - \dot{A} &= 0 , \label{eq:0iv2} \\
    -\Ddot{A} - (3H + \Gamma) \dot{A} + H \dot{E} + \frac{3}{2} H^2 E &= 0 , \label{eq:iiv2} \\
    -A - E + 2\dot{\sigma} + 2(H + \Gamma) \sigma &= 0 \,. \label{eq:ijv2}
\end{align}
This equation can be combined to show that only one of the three scalar fields $E$, $A$, and $\sigma$ is dynamical as desired for a single clock cosmology.

The Bardeen potentials are $\Phi = 1/2 E - \dot{\sigma}$ and $\Psi = -1/2 A + H \sigma$. Solving the $00$ and $0i$ equations \eqref{eq:00v2} and \eqref{eq:0iv2} respectively, we obtain expressions for $\sigma$ and $E$ in terms of $A$ and $\dot{A}$. Combining them with the transverse equation \eqref{eq:ijv2} yields a second order differential equation for $A$, from which we deduce
\begin{align}\label{eq:PsiEOM}
    f_\Psi \left( \ddot{\Psi} + \frac{c_s^2 k^2}{a^2} \Psi  + \nu_{\Psi} \dot{\Psi} + \mu_{\Psi} \Psi \right) = 0,
\end{align}
with $f_\Psi \equiv (3H + 4\Gamma)$, the speed of sound 
\begin{align}
    c_s^2 \equiv \frac{H f_\Psi}{3H^2 + 2M^2}\,,
\end{align}
and
\begin{align}
    f_\Psi \cdot \nu_{\Psi} &\equiv 12H^2 + 19H \Gamma + 8 \Gamma^2 - 4\dot{\Gamma} \,, \\
    f_\Psi \cdot \mu_{\Psi} &\equiv - 2 H( 3H \Gamma + 2\Gamma^2 + 2\dot{\Gamma} )\,.
\end{align}
This equation of motion contains four physical ingredients: a propagation speed $c_s$, a friction term $\nu_\Psi$, an effective mass $\mu_\Psi$, and a kinetic prefactor $f_\Psi$ setting the normalization of the kinetic terms. From now on, we take the limit $M \to \infty$, corresponding to a vanishing sound speed $c_s \to 0$ for the propagating scalar mode, thereby recovering the pure dust universe when $\Gamma = 0$. 
In the limit $\Gamma \to -3H/4$, the kinetic coefficient satisfies $f_\Psi \to 0$, so that the kinetic terms vanish while the friction and mass terms remain finite. The second-order equation then reduces to a first-order evolution equation. Since this occurs in an attractor regime where $|\ddot{\Psi}| \ll |\nu_\Psi \dot{\Psi}|,\, |\mu_\Psi \Psi|$, the transition is smooth and does not introduce any discontinuity.

\begin{figure}[t]
    \includegraphics[width=1\linewidth]{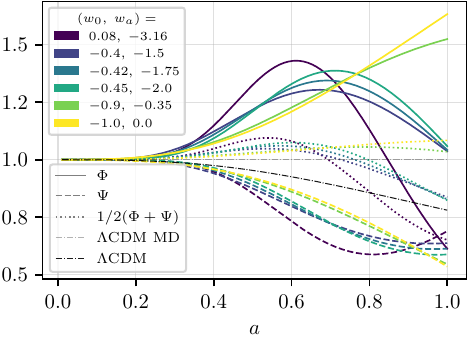}
    \caption{Bardeen potentials $\Phi$ and $\Psi$ predicted by our model (ODF), along with their average, and the corresponding evolution in $\Lambda$CDM and a matter-dominated universe. All fields are normalized to unity at the initial time.}
    \label{fig:pot}
\end{figure}

\begin{figure}[t]
    \includegraphics[width=1\linewidth]{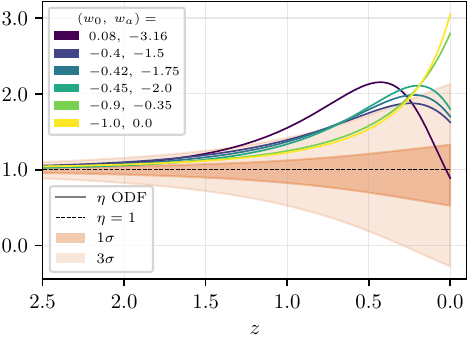}
    \caption{Gravitational slip $\eta(z)=\Phi(z)/\Psi(z)$ predicted by our model (ODF), versus BAO+CMB constraints \cite{DESI:2024hhd}. The dashed line indicates $\Lambda$CDM, where $\eta=1$.}
    \label{fig:eta}
\end{figure}

Finally, $\Phi$ is expressed in terms of $\Psi$ and $\dot{\Psi}$ using the expressions for $E$ and $\sigma$ derived above. Equation~\eqref{eq:PsiEOM} is then solved numerically for the choices of $\Gamma$ given in Eq.~\eqref{eq:gamma}. The resulting gravitational potentials are shown in Fig.~\ref{fig:pot} for different functions $\Gamma(w_0, w_a)$. Compared to $\Lambda$CDM, $\Phi$ is typically enhanced while $\Psi$ is suppressed. Interestingly, their sum $\Phi+\Psi$, which determines weak lensing, remains closer to its $\Lambda$CDM counterpart.
Expressing $\Phi$ in terms of $\Psi$ and $\dot{\Psi}$ yields a differential equation for the gravitational slip $\eta \equiv \Phi/\Psi$ given in the Supplementary Material. It can be solved analytically to give
\begin{align}\label{eq:eta_formula}
    \eta = 1 + \frac{\Gamma}{H}\Bigg\{&1 - a^{1/2} H^2(a) \Bigg[ \frac{1}{a^{1/2}_i H^2(a_i)} \nonumber \\
    +& \int_{a_i}^{a} \dd \tilde{a} \, \frac{3H(\tilde{a}) + 4\Gamma(\tilde{a})}{2 \tilde{a}^{3/2} H^3(\tilde{a})} \Bigg]^{-1}\Bigg\}\,,
\end{align}
which recovers $\eta(a_i) = 1$ at initial time. 
A departure from $\eta = 1$ signals a deviation from general relativity and is tightly constrained by large-scale structure observations, including DESI. Fig.~\ref{fig:eta} displays the current bounds, showing that the present parametrization lies near the boundary of the $3\sigma$ contour. For a more accurate comparison with data, one should include baryons in our analysis and perform a scan over cosmological parameters.

The last observable we consider is galaxy clustering, through the evolution of baryons, which has so far been neglected. In contrast to the dark sector, whose conservation is non-standard as in Eq.~\eqref{eq:continuity}, we assume that baryons are conserved in the usual way, $\nabla^\mu T^{(\rm b)}_{\mu\nu}=0$. This condition leads to the usual continuity and Euler equations, which combine into the evolution equation for the baryon density contrast $\delta_{\rm b} \equiv \delta\rho_{\rm b}/\bar{\rho}_{\rm b}$ in the Newtonian gauge
\begin{equation}\label{eq:growth_of_structure}
\ddot{\delta}_{\rm b} + 2H \dot{\delta}_{\rm b} = - \frac{k^2}{a^2} \Phi + 3 \ddot{\Psi} + 6 H \dot{\Psi}\,.
\end{equation}
In the the sub-Hubble limit, time derivatives of $\Psi$ are small compared to spatial gradients and the evolution of $\delta_{\rm b}$ is sourced using our solution for $\Phi$. The modified dynamics does not introduce scale dependence, so that the usual ansatz $\delta_{\rm b}(a, \bfk)= D(a) \cdot \delta_{\bfk, 0}$ remains valid, separating the growth factor $D(a)$ from the adiabatic initial conditions $\delta_{\bfk, 0}$, and similarly for $\Phi (a, \bfk) = \Phi(a) \cdot \Phi_{\bfk, 0} $. Details of the numerical implementation can be found in the Supplementary Material. 

\begin{figure}[t]
    \includegraphics[width=1\linewidth]{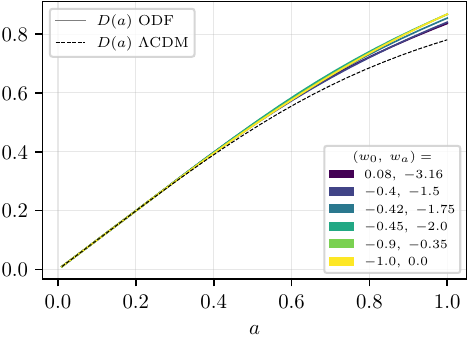}
    \caption{Growth factor $D(a)$ in our model (ODF) compared to $\Lambda$CDM.}
    \label{fig:growth_rate}
\end{figure}

Solutions for the growth factor $D(a)$ are shown in Fig.~\ref{fig:growth_rate}. At early times, $D(a)\propto a$, as in a standard matter-dominated universe. At late times, the growth rate interpolates between the matter-dominated behavior and that of $\Lambda$CDM, leading to enhanced structure formation relative to the latter. As a result, the present-day amplitude of fluctuations is increased, with $\sigma_8(a_0=1)$ larger than its $\Lambda$CDM value by a factor $D_{\mathrm{ODF}}(a_0)/D_{\Lambda\mathrm{-CDM}}(a_0)\sim [1.07,1.11]$. This enhancement places the minimal scenario considered here outside the observationally allowed range. More precise bounds would require scanning over cosmological parameters, a precise inclusion of baryonic effects and the consideration of stochastic sources.  

\begin{figure}[t]
    \includegraphics[width=1\linewidth]{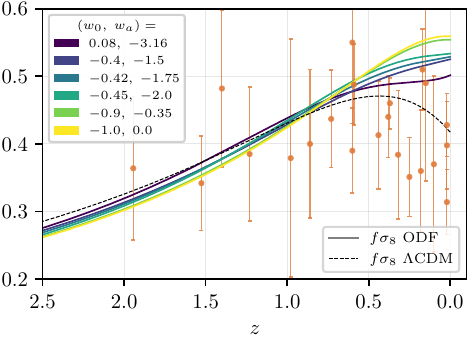}
    \caption{Predicted $f \sigma_8$ in our model (ODF) and $\Lambda$CDM, compared to observational data points from \cite{Sagredo:2018ahx}.}
     \label{fig:fsigma8}
\end{figure}

A related quantity tightly constrained by observations is $f\sigma_8$, where $f(a)=\dd\ln D(a)/\dd\ln a$ measures the time variation of structure growth. Figure~\ref{fig:fsigma8} shows this observable at low redshift together with observational constraints from \cite{Sagredo:2018ahx}. The model again predicts an excess of structure formation over a range of scales. Whether additional physical effects could compensate for this enhancement is left for future work. 
\\

\paragraph*{Outlook.---} 

We have proposed and studied a unified description of the dark sector based on an open-system formulation of gravity. The resulting dynamics fits the expansion history as measured by CMB and BAOs, while predicting distinctive signatures in the gravitational-wave luminosity distance, galaxy clustering, and weak lensing.
This framework provides a natural embedding for single-clock models of an interacting dark-sector, and will be tested by next-generation surveys including DESI and Euclid. Incorporating stochastic noise and other operators may further affect structure formation, and the resulting dynamics could leave observable imprints in the integrated Sachs–Wolfe effect.
Our results predict concrete, model-agnostic signatures for upcoming precision probes of cosmic acceleration.\\

\paragraph*{Acknowledgments.---} 
We thank Tessa Baker, Camille Bonvin, Pedro Ferreira, Mariana Carrillo Gonzalez, Austin Joyce, Wayne Hu, Frank Qu, Antony Lewis, Maria Mylova, Toshifumi Noumi, Alessandra Silvestri and Sarunas Verner for the insightful discussions. This work has been supported by STFC consolidated grant ST/X001113/1, ST/T000694/1, ST/X000664/1, ST/Y509127/1 and EP/V048422/1. S.A.S. was supported by a Harding Distinguished Postgraduate Scholarship. This work is supported by the Kavli Institute for Cosmological Physics at the University of Chicago, and the Kavli Foundation.

\bibliographystyle{JHEP}
\bibliography{biblio}

\clearpage

%%%%%%%%%%%%%%%%%%%%%%%%%%%%%%
\clearpage
\appendix

\begin{widetext}

\section{Supplementary Material}

This supplementary material contains some technical details of the derivations presented in the main text. Based on the perturbed Einstein equations (\ref{eq:00v2}-\ref{eq:ijv2}), we derive a differential equation obeyed by the gravitational slip. We then discuss the derivation of the growth function for baryons. 

\subsubsection{Gravitational slip}

The gravitational slip $\eta$ quantifies deviations from $\Phi=\Psi$ through
\begin{equation}\label{eq:appeta}
\eta=\frac{\Phi}{\Psi}= \frac{(3 H+2 \Gamma )-2 \frac{\Gamma}{H}  \frac{\dot{\Psi}}{\Psi}}{(3 H+4 \Gamma )}\,,
\end{equation}
where $\eta$ depends explicitly on $\dot{\Psi}/\Psi$. We address this issue by finding the differential equation that governs $\eta$. To derive this equation, we take the time derivative of \eqref{eq:appeta} and substitute the equation of motion for $\Psi$ \eqref{eq:growth_of_structure} to remove $\ddot{\Psi}$. Finally, we can remove any dependence on the ration $\dot{\Psi}/\Psi$ using \eqref{eq:appeta}. This leaves us with
\begin{equation}\label{eq:appetaEOM}
  \dot{\eta} =  \frac{3 H^2 (\eta -1)^2}{2 \Gamma }+H \left(2 \eta ^2-3 \eta +1\right)+\frac{(\eta -1) \dot{\Gamma}}{\Gamma }-\Gamma  \eta \,. 
\end{equation}
If for some time $t_{i}$ we have that $\eta\left(t_{i}\right)=1$ then $\dot{\eta}\left(t\right)|_{t=t_{i}}=-\Gamma(t)|_{t=t_{i}}$, which reflects that any deviation from $\Phi=\Psi$ is induced by the effects of the dissipation operator. It is also possible to find an analytic solution for $\eta$ given by the main text result \eqref{eq:eta_formula}. 
    
\subsubsection{Galaxy clustering}

Here, we study the baryon clustering in the model presented in this letter. For this, we follow the time evolution of the density contrast of a tracer fluid that has zero pressure and zero speed of sound
\begin{equation}
T^{\rm b}_{\mu\nu}=\rho_{\rm b}\left(t,\textbf{x}\right)u^{\rm b}_{\mu}u^{\rm b}_{\nu} \,,\quad\mathrm{s.t.}\quad g^{\mu\nu}u^{\rm b}_{\mu}u^{\rm b}_{\nu}=-1\,.
\end{equation}
In this work, we neglect its contribution to the Einstein equations, considered subdominant. Hence, its dynamics is fully characterized through its conservation equation $\nabla^{\mu}T^{\rm b}_{\mu\nu}=0$, which we assume to be standard. The conservation equation leads to the continuity and Euler equations
\begin{equation} \label{eq:appconservation}
    \dot{\delta}_{\rm b} + \frac{\nabla^2}{a^2} v_{\rm b} - 3 \dot{\psi}_N = 0\, \quad \mathrm{and} \quad \dot{v}_{\rm b} + \phi_N = 0\,,
\end{equation}
where, in the Newtonian gauge $\phi_N = \Phi$ and $\psi_N = \Psi$. The Euler and continuity equations combine into the main text result Eq.~\eqref{eq:growth_of_structure}. 

The time evolution of the density contrast does not take a factorized form in general, yet, on sub-Hubble scales, the time derivatives of the Bardeen potentials in the last two terms of Eq.~\eqref{eq:growth_of_structure} become negligible compared to the spatial gradient, leading to 
\begin{equation}
    \ddot{\delta}_{\rm b} + 2H \dot{\delta}_{\rm b} \simeq - \frac{k^2}{a^2} \Phi .
\end{equation}
In this limit, the density contrast can be expressed in terms of the growth rate $D(a)$ through
\begin{align}
    \delta_{\rm b}\left(a, \bfk\right)&=\delta_{\bfk, 0}  \cdot D(a)\,,
\end{align}
where $D(a_i) = a_i$ and $D'(a_i) = 1$ at initial time $a_{i}\ll1$, just as the standard $\Lambda$CDM case. Similarly, the Bardeen potential $\Phi$ obeys an equation of motion alike Eq.~\eqref{eq:PsiEOM}, obtained by using Eq.~\eqref{eq:eta_formula}. These equations are momentum independent, such that we can write their evolution in a factorized form
\begin{align}
    \Phi\left(a, \bfk\right) &=\Phi_{\bfk, 0} \cdot  \Phi(a)\,.
\end{align}
We are left with an equation for $D(a)$ of the form
\begin{equation}\label{eq:growthrateapp}
     a^2 \frac{\dd^2 D(a)}{\dd a^2} + a \left(3 +  a \frac{\dd \ln H}{\dd a} \right)\frac{\dd D(a)}{\dd a} + \frac{k^2}{a^2 H(a)^2 } \frac{\Phi_{\bfk, 0}}{\delta_{\bfk, 0}}\Phi(a) = 0,
\end{equation}
where the ratio between the initial condition for the Bardeen potential and the density contrast of the tracer field can be inferred from the early time limit of the Einstein equation. At early times, the contribution of $\Gamma$ is negligible and we reduce to the standard matter dominated universe. In particular, at early times and on sub-Hubble scales, the Poisson equation applies, leading to
\begin{equation}\label{eq:Poisson}
    - \frac{k^2}{a^{2}_i}  \Phi_{\bfk, 0} \cdot \Phi(a_i) =  \frac{3}{2} H^2(a_i)D(a_i) \delta_{\bfk, 0} \,,
\end{equation}
which fixes the ratio $\Phi_{\bfk, 0}/\delta_{\bfk, 0}$. Combining \eqref{eq:growthrateapp} and \eqref{eq:Poisson}, we can use numerical integration to produce Fig.~\ref{fig:growth_rate}.

\end{widetext}

\end{document}